\documentclass{llncs}
\usepackage{llncsdoc}
\usepackage{graphicx}
\usepackage{hyperref}

\begin{document}

\title{
Practical
Implementation of Lattice QCD Simulation
on SIMD Machines with Intel AVX-512\thanks{%
The final authorized version is avaiable online at \url{https://doi.org/10.1007/978-3-319-95168-3_31}}
}

\author{Issaku Kanamori\inst{1} \and Hideo Matsufuru\inst{2}}

\institute{
Department of Physical Science, Hiroshima University, \\
Higashi-hiroshima 739-8526, Japan\\
\email{kanamori@hiroshima-u.ac.jp}
\and 
High Energy Accelerator Research Organization (KEK),\\
1-1 Oho, Tsukuba, Ibaraki 305-0801, Japan\\
\email{hideo.matsufuru@kek.jp}
}

\maketitle

\begin{abstract}
We investigate implementation of lattice Quantum Chromodynamics (QCD)
code on the Intel AVX-512 architecture.
The most time consuming part of the numerical simulations of
lattice QCD is a solver of linear equation for a large sparse matrix
that represents the strong interaction among quarks.
To establish widely applicable prescriptions, we examine rather
general methods for the SIMD architecture of AVX-512, such as using
 intrinsics and manual prefetching, for the matrix multiplication.
Based on experience on the Oakforest-PACS system,
a large scale cluster composed of Intel Xeon Phi Knights Landing,
we discuss the performance tuning exploiting AVX-512 and code design
on the SIMD architecture and massively parallel machines.
We observe that the same code runs efficiently on an
Intel Xeon Skylake-SP machine.
\end{abstract}

\section{Introduction}
\label{sec:Introduction}

Quantum Chromodynamics (QCD) is the fundamental theory to describe
the strong interaction among quarks.
QCD plays a crucial role not only in understanding the properties of
nucleons but also in seeking for new fundamental physics behind the
background of QCD through precise theoretical calculation.
Because of its nonlinear nature and large coupling at low energy,
however, QCD is generally not solved even though the fundamental
equation is known.
Lattice QCD, which formulates QCD on Euclidean 4-dimensional
spacetime lattice, provides a numerical method to tackle this
problem \cite{textbook}.
Based on the path integral formulation of quantum field theory,
the partition function becomes similar to that of the statistical
mechanics so that the Monte Carlo methods are applicable.
Typically the most time consuming part of the lattice QCD
simulations is solving a linear equation for a large sparse fermion
matrix that represents the interaction among quarks.
The numerical cost grows rapidly as increasing the lattice size
toward precision calculation.
Thus the lattice QCD simulations have been a typical problem
in high performance computing and progressed keeping step
with development of supercomputers.

One of the trends in high performance computer architecture is to possess
long SIMD vector registers.
The Intel Xeon Phi series is the first product of Intel that has
a SIMD vector of 512-bit length.
Its second generation,
the Knights Landing (KNL) adopts the AVX-512 instruction set.
This new instruction set is now available also on the
Intel Xeon Skylake-SP series.

In this paper, we port a lattice QCD simulation code to
machines composed of KNL and Skylake-SP processors.
The aim of this work is to establish techniques to exploit
AVX-512, or more generally SIMD architecture, that are
applicable to wide range of applications.
Since one frequently needs to port a legacy code to a new
architecture, it is important to acquire such simple prescriptions
to achieve acceptable (not necessarily the best) performance.
It is also helpful for designing code structure in future
development.
For this reason, we restrict ourselves in rather general
prescriptions: change of data layout, application of Intel AVX-512
intrinsics, and prefetching.
As a testbed of our analysis, we choose two types of fermion
matrices together with an iterative linear equation solver.
In our previous report \cite{Kanamori:2017tlp,Kanamori:2017urm},
we developed a code along the above policy and applied it to KNL.
In this paper, in addition to improved performance, we rearrange
these prescriptions so that each effect is more apparent.
As for the fermion matrices, one of those adopted in
\cite{Kanamori:2017tlp,Kanamori:2017urm} is replaced with other
widely used matrix, aiming at extending application of the
developed techniques.
As a new target machine, we examine a cluster system composed
of the Intel Skylake-SP processor.

Here we briefly summarize related works.
Since the Skylake-SP is rather new, works related to lattice QCD
are so far for KNL.
In a KNL textbook \cite{KNLtextbook}, Chapter 26 is devoted to
performance tuning of
the Wilson-Dslash operator (one of matrices examined in this work)
using the QPhiX library \cite{QPhiX}.
Ref.~\cite{Boyle:2016lbp} developed a library named `Grid'
for the SIMD architecture, which has largely affected our work.
In Ref.~\cite{Boyle:2017xcy}, through the discussion of memory page
granularity, Grid is examined on the Skylake-SP system.
A quite large scale test with a preconditioned clover-type fermion
matrix is reported in \cite{Boku:2017urp}.
A pragma-based recipe is examined in \cite{Durr:2017clx}.
A plenary review at the annual lattice conference \cite{Rago:2017pyb}
summarizes recent activities in lattice QCD community.

This paper is organized as follows.
The next section briefly introduces the linear equation system
in lattice QCD simulations with fermion matrices employed
in this work.
Features of the AVX-512 instruction set and our target architectures
are summarized in
Section~\ref{sec:AVX-512}.
Section~\ref{sec:implementation} describes the details of
our implementation.
We examine the performance of our code on the systems composed of KNL
and Skylake in Sections~\ref{sec:KNL} and \ref{eq:Skylake}, respectively.
We conclude by discussing implication of our results in the last
section.

\section{Lattice QCD Simulation}
\label{sec:lattice_QCD}

For the formulation of lattice QCD and the principle of numerical
simulation, there are many textbooks and reviews \cite{textbook}.
Thus we concentrate on the linear equation for the fermion matrix
to which high computational costs are required.

The lattice QCD theory consists of fermion (quark) fields
and a gauge (gluon) field.
The latter mediates interaction among quarks and are represented
by `link variable', $U_\mu(x)\in SU(3)$, where
$x=(x_1, x_2, x_3, x_4)$ stands for a lattice site and
$\mu=1,2,3,4$ is the spacetime direction.
In numerical simulations the lattice size is finite:
$x_\mu=1,2,\dots, L_\mu$.
The fermion field is represented as a complex vector on
lattice sites,
which carries 3 components of `color' and 4 components of `spinor',
thus in total 12,
degrees of freedom on each site.
The dynamics of fermion is governed by a functional 
$S_F=\sum_{x,y} \psi^\dag(x) D[U]^{-1}(x,y) \psi(y)$,
where $D[U]$ is a fermion matrix acting on a fermion vector $\psi(x)$.
A Monte Carlo algorithm is applied to generate an ensemble
of the gauge field $\{U_\mu(x) \}$, that requires to solve
a linear equation $x = D^{-1} \psi$ many times.

There is a variety of the fermion operator $D[U]$, since its
requirement is to coincide with that of QCD in the continuum limit,
the lattice spacing $a\rightarrow 0$.
Each formulation has advantages and disadvantages.
As a common feature, the matrix is sparse because of the locality
of the interaction.
In this paper, we examine the following two types of fermion matrix.

\subsubsection{Wilson Fermion Matrix}
\begin{figure}[tb]
\centering
\includegraphics[width=4.1cm]{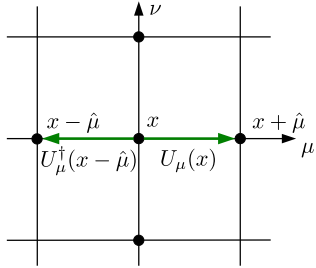}
\vspace{-1mm}
\caption{
The schematic feature of the Wilson fermion matrix}
\label{fig:Wilson_fermion_matrix}
\end{figure}

The first one called the Wilson fermion matrix has the form
\begin{equation}
 D_{\mathrm{W}}(x,y) = \delta_{x,y} - \kappa \sum_{\mu=1}^4
           \big[ (1-\gamma_\mu) U_\mu(x) \delta_{x+\hat{\mu},y}
        + (1+\gamma_\mu) U_\mu^\dag(x-\hat{\mu}) \delta_{x-\hat{\mu},y}
         \big] ,
\label{eq:Wilson_fermion_operator}
\end{equation}
\noindent
where $x$, $y$ are lattice sites,
$\hat{\mu}$ the unit vector along $\mu$-th axis,
and $\kappa=1/(8+2m_0)$ a parameter related to the quark mass $m_0$.
Fig.~\ref{fig:Wilson_fermion_matrix} indicates how the interaction
to the neighboring sites is involved in the matrix.
As mentioned above, the link variable $U_\mu(x)$ is 
a $3\times 3$ complex matrix acting on the color
and $\gamma_\mu$ is a $4\times 4$
matrix acting on the spinor degrees of freedom.
Thus 
$D_{\mathrm{W}}$ is a complex matrix of the rank $4 N_c L_x L_y L_z L_t$.
It is standard to impose the periodic or anti-periodic boundary
conditions.

\subsubsection{Clover Fermion Matrix}
The second fermion matrix called `clover' fermion is an improved
version of the Wilson fermion so as to reduce the discretization error.
It is widely used in practical simulation due to its moderate
numerical cost.
The clover fermion matrix is defined as
\begin{equation}
 D_{\mathrm{clov}}(x,y)= D_{\mathrm{W}}(x,y) + F(x)\delta_{x,y},
\end{equation}
where $F(x)$ in the additional term is a $12\times 12$
Hermitian matrix made of link variables.
By choosing a proper basis, $F(x)$ is represented as a block diagonal
form with two $6\times 6$ Hermitian matrices.
Note that $F(x)$ is determined before the solver algorithm is
applied and stored as an array data similarly to the link variable
$U_\mu(x)$ so that its building cost is ignored in this paper.

\begin{table}[t]
\caption{
Features of the fermion matrices:
The number of floating point operations and the data transfer
between memory and processor per site.
The so-called roofline estimates for the Oakforest-PACS (roofline 1)
and ITO (roofline 2) are obtained from the values of Byte/Flop.
The ideal cases assume that the data are loaded from the memory
only once.}
\begin{tabular}{lccccc}
\hline\noalign{\smallskip}
 Fermion type &
 $N_{\textrm{flop}}$/site &
 \parbox{4.5em}{data/site\\ float [B] } &
 Byte/Flop &
 \parbox{8em}{roofline 1 (ideal)\\ \hspace*{3em} [GFlops]} &
     \parbox{8em}{roofline 2
     (ideal)\\ \hspace*{3em} [GFlops]}\\
 \noalign{\smallskip}
\hline
\noalign{\smallskip}
 Wilson       & 1368  & 1536  & 1.12  & 424 (1350) & 228 (729)
		      \\
 Clover       & 1944 & 1824  & 0.94  & 506 (1200)  & 273 (648) \\
 \hline
\end{tabular}
\label{comparison_of_fermion_matrix}
\end{table}

The above two fermion matrices have differences in data size
transferred between the memory and the processor cores, and number of
arithmetic operations.
Table~\ref{comparison_of_fermion_matrix} summarizes these values
per site for single precision data.
As quoted in Table~\ref{comparison_of_fermion_matrix},
the clover matrix tends to have smaller byte-per-flop
value, due to the multiplication of $F(x)$.
With caches one can expect reuse of data.
Every link variable $U_\mu(x)$ and vector on each site
can be reused maximally two and nine times, respectively.
In the ideal case, byte-per-flop ratio becomes 0.351 and 0.395 for
the Wilson and clover matrices, respectively.
The ratio becomes larger for the clover matrix since $F(x)$ is used only once.
The last two columns in Table~\ref{comparison_of_fermion_matrix}
are so-called roofline estimates,
the performance limit determined by the memory bandwidth,
for our target machines, Oakforest-PACS and ITO.
For the former, the bandwidth of the MCDRAM is used.
The ideal roofline is estimated by assuming that the data are loaded
from the memory only once and successive access is to the cache.

As mentioned above, we need to solve a linear equation system with
the fermion matrix.
Since the fermion matrices are large and sparse,
iterative solvers based on the Krylov subspace method are
widely used.
We employ the BiCGStab algorithm
for a non-Hermitian matrix.
We focus on the performance with single precision,
since it is practically used as an inner solver 
of a mixed precision solver and governs the performance of
the whole solver.
While there are variety of improved solver algorithms for
a large-scale linear systems, such as a multi-grid or
domain-decomposition methods,
they are beyond the scope of this paper.

\section{Target Architectures}
\label{sec:AVX-512}

Our target architectures adopt the Intel AVX-512 instruction set.
Each thread can use 32 vector registers of 512-bit length which
can process 16 single or 8 double precision floating point
numbers simultaneously.
A fused multiplication and add (FMA) operation on SIMD registers
can perform 32 FLOPs (single precision) and 16 FLOPs (double).
The full instruction set consists of several sub-categories
among which all the processor launched so far can exploit AVX-512F
(Foundation) and AVX-512CD (Conflict Detection).

\subsubsection{Knights Landing}
The first example with AVX-512 is 
the Intel Xeon Phi Knights Landing (KNL).
It is the second generation of Intel Xeon Phi
architecture, whose details are found in \cite{KNLtextbook}.
It is composed of maximally 72 cores, in units of a tile
that is composed of two cores sharing distributed L2 cache.
Each core supports 4-way hyper-threading.
In addition to DDR4 memory with about 90 GB/s, MCDRAM of maximally 16 GB
accessible with 475--490 GB/s \cite{KNLmemory}
is available with one of three modes: cache, flat, and hybrid.
Our target machine with KNL is the Oakforest-PACS system hosted by
Joint Center for Advances High Performance Computing
(JCAHPC, University of Tokyo and University of Tsukuba).
The system is composed of 8208 nodes of Intel Xeon Phi
7250 (68 cores, 1.4 GHz) connected by full-bisection fat tree
network of the Intel Omni-Path interconnect.
It has 25 PFlops of peak performance in total, and started
public service in April 2017.

\subsubsection{Skylake-SP}
The AVX-512 instruction set has become available on
the new generation of Intel Xeon processor, Skylake-SP.
We use a system named ITO at 
the Research Institute for Information Technology, Kyushu University.
It started full operation in January 2018.
Each node of ITO has two Intel Xeon Gold 6154 (Skylake-SP, 18 cores, 3.0
GHz) processors
which amounts to 6.9 TFlops/node of peak performance in single precision
(slightly larger than single KNL node).
The main difference from KNL is the memory structure: instead of
MCDRAM, Skylake-SP has a L3 cache. It is shared by all cores in
CPU and its size is 24.75MB.
The bandwidth of the main memory (DDR4, 192 GB/node) is 255.9
GB/s/node.

\section{Implementation}
\label{sec:implementation}

Our base code is the Bridge++ code set
\cite{bridge_website,Ueda:2014zsa} which is
described in C++ based on an object oriented design
and allows us to replace fermion matrices and solver algorithms
independently.
In the original Bridge++, hereafter called the
Bridge++ core library, the data are in double precision and
in a fixed data layout with array of structure (AoS).
Following the strategy employed in
\cite{Motoki:iccs2014},
we extend Bridge++ to enable a flexible data layout and an arbitrary precision
type.
The key ingredients of our implementation are as follows:
changing data layout, use of intrinsics, manual prefetching,
assignment of the thread tasks. 
In the following, these issues are described in order in some detail.
 
\subsubsection{Data Layout}
It is important to choose a proper data layout to attain high affinity
to the SIMD vector registers.
As a 512-bit register corresponds to 8 or 16 floating point
numbers in double and single precision, respectively,
we rearrange the date in these units.
We implement the code in C++ template classes and instantiate
them for the double and float data types individually.
There are several ways in ordering the real and imaginary parts
of complex variables.
Considering the number of SIMD registers and the number of
the degree of freedom on each site, we decide to place
the real and imaginary parts as consecutive data on the memory.
The color and spinor components are distributed to separate
registers.
Instead several sites are packed into a SIMD vector;
complex variables of float (double) type on eight (four)
sites are processed simultaneously.
To allocate the data on the memory, we use \texttt{std::vector}
in the standard C++ template library with providing an aligned allocator.

\begin{figure}[t]
\centering
 \includegraphics[scale=0.5]{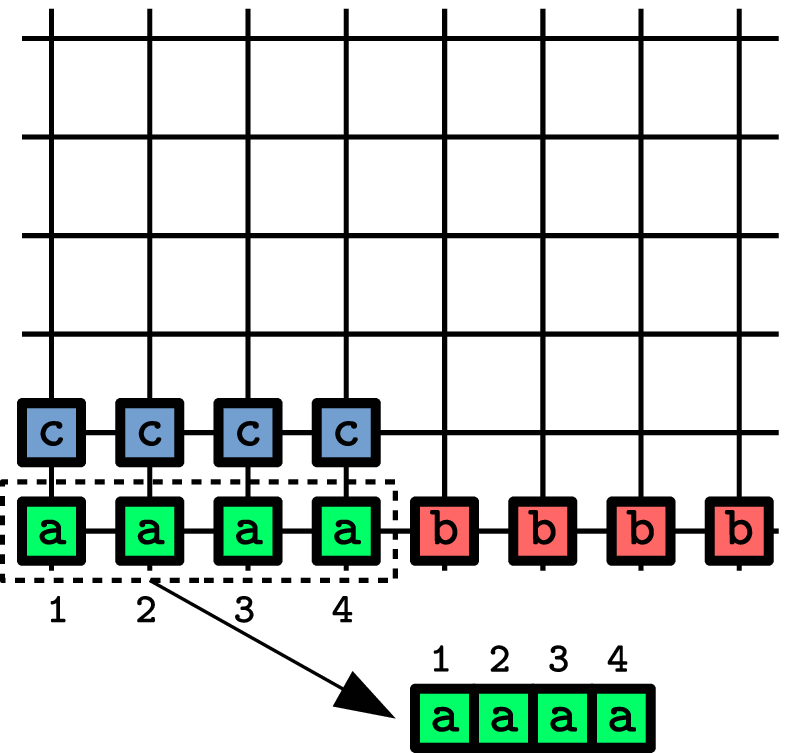}
\hfil
 \includegraphics[scale=0.5]{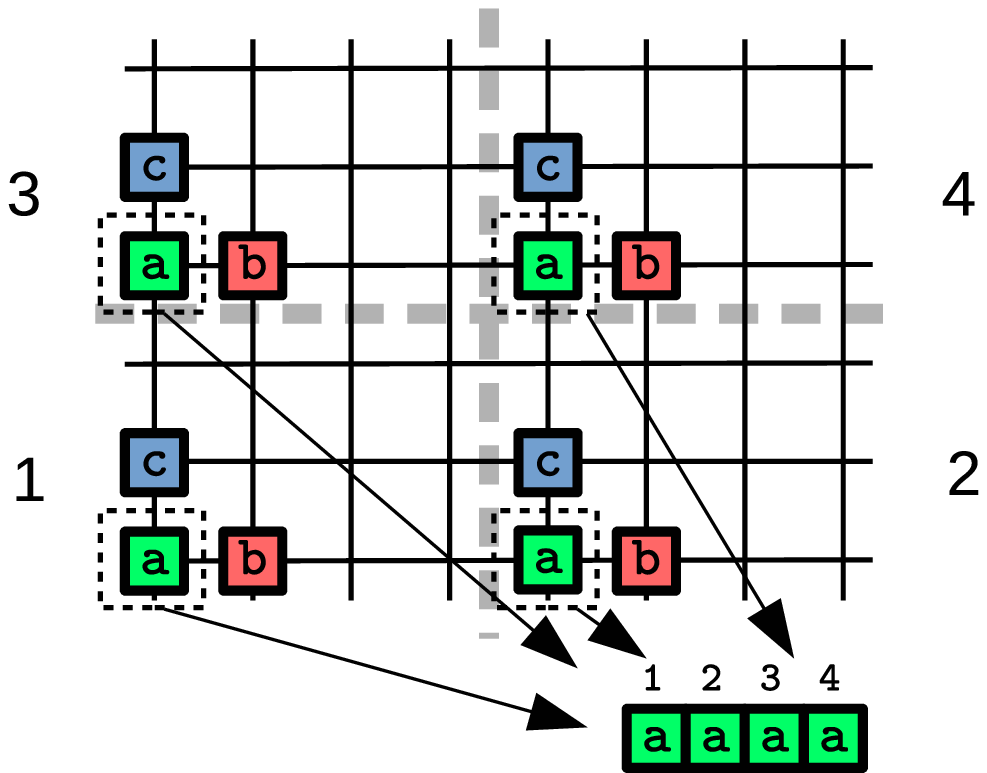}
\caption{
The site index ordering for the layout 1 (left) and 2 (right)
in the double precision case.
We use a three-dimensional analogues of the right panel for
the layout 2 in single precision.}
\label{fig:layout_site_ordering}
\end{figure}

There is still flexibility in folding the lattice sites into
a data array.
We compare two different data layouts displayed
in Fig.~\ref{fig:layout_site_ordering}.
To avoid lengthy description, we assume the single precision
case in the following.
In the first case (layout 1), several sites in $x$-coordinate
composes a SIMD vector.
This requires the local lattice size in $x$-direction to be a
multiple of eight.
Since the $x$-coordinate is the most inner coordinate of our
index, it is a simplest extension of a non-vectorized layout.
To minimize performance penalty of boundary copy, the MPI
parallelization is not applied in $x$-direction.

The second layout (layout 2) was introduced in
Ref.~\cite{Boyle:2016lbp}.
As the right panel of Fig.~\ref{fig:layout_site_ordering}
explains, the local lattice is divided into several subdomains
each provides one complex number to one SIMD vector.
With our implementation this restricts the local lattice sizes
in $y$-, $z$-, and $t$-directions to be even.
While there is no restriction in $x$-direction for layout 2,
throughout this paper we do not MPI parallelize in $x$-direction
similarly to the layout 1.

\subsubsection{Using Intrinsics}

The arithmetic operations on the SIMD variables are
explicitly managed using the intrinsics.
We wrap them in inline functions, which cover common basic
operations such as complex four arithmetic operations and
BLAS-like functions (axpy, etc.).
By replacing the wrapper function, 
our code can be easily adapted to other architectures
such as AVX2.
We partially make use of the
\texttt{simd} directory in the Grid library \cite{Boyle:2016lbp}
that provides similar wrappers.
Defining types that wrap the 512-bit SIMD vectors
(\texttt{\_\_m512} or \texttt{\_\_m512d}) and using arrays of them,
the compiler generates a load/store instruction for a vector register.
We therefore do not explicitly use load or store intrinsics,
except for streaming stores.

\subsubsection{Prefetching}

We compare the manual prefetch and the automated prefetch by
compiler.
The most outer loop of the matrix is for the site index (modulo
SIMD vector).
At each site, one accumulates nine stencil contributions, from $+x$,
$-x$,...,$-t$ directions in order, and of that site.
It turned out that in most cases only the prefetch to L2 cache is
relevant.
Manual prefetch to L1 cache sometimes causes slowing down%
\footnote{
Removal of
such prefetch is one of the reasons of improved performance from
our previous report \cite{Kanamori:2017urm}.
}.
The prefetch to L2 cache is inserted at three steps before the
computation except for the contributions from the $-x$ direction
and that site.
Since $x$ is the innermost site index, 
the neighboring data in $-x$ direction most likely remain
in the cache.
For the clover fermion matrix, additionally two $6\times 6$ block
matrices in the clover term $F(x)$ are multiplied.
The prefetch is applied only to the first block matrix.
To trigger the hardware prefetch, a few more prefetches
to L1 cache are inserted: (i) at the beginning of the site loop
in order to load the data in $+x$ direction,
and (ii) at the almost end of multiplication of the first block
matrix of the clover term to load the second block matrix.
We also insert several L2 and L1 prefetches during the packing
of data for communication and index calculation so as to load
the local lattice sizes.

We use \texttt{\_mm\_prefetch} with
\texttt{\_MM\_HINT\_T0} and \texttt{\_MM\_HINT\_T1}
to generate the prefetch order.
The following pseudo-code is an example of
prefetch insertions.
{\small
\begin{verbatim}
for(s=0; s<num_of_sites; s++){
#pragma noprefetch
  // +x
  prefetch_to_L1(+x);
  prefetch_to_L2(-y);
  accumulate_from(+x);
  // -x
  prefetch_to_L2(+z);
  accumulate_from(-x);
  // +y
  prefetch_to_L2(-z);
  accumulate_from(+y);

  ...
}
\end{verbatim}
}
\noindent
It is not straightforward to insert prefetch commands
appropriately.
One needs to tune the variables and the place to insert
referring to a profiler, {\it e.g.} Intel Vtune amplifier.
The performance may be sensitive to the problem size,
choice of parameters such as the number of threads,
and so on.

\subsubsection{Thread Task Assignment}

Since the lattice usually extends over at least several nodes, 
a multiplication of matrix requires communication among nodes.
The matrix multiplication has the following steps in order:
(1) Packing of the boundary data for communication,
(2-a) Doing communication,
(2-b) Operations of the bulk part, and
(3) Unpacking the boundary data and operations on the boundary part.
(2-a) and (2-b) can be overlapped, and its efficiency is the 
key for the total performance.
We restrict ourselves in the case that only the master thread
performs the communication, {\it i.e.} corresponding to
{\tt MPI\_THREAD\_FUNNELED}.
For the implementation of the steps (2-a) and (2-b) above,
there are two possibilities:
(i) arithmetic operational tasks are equally assigned to all
the available threads, and
(ii) the master thread concentrates the communication and
other threads bear the arithmetic tasks.
We adopt the case (ii) in this work, since it tends to be faster. 

In order to make the extrapolation of the scaling easier,
we always enforce the above communication procedure
in all the $y$-, $z$- and $t$-directions even if no MPI
parallelization is imposed, so that the ``communication''
may just result in a copy of packed data.

\section{Performance on KNL Machine: Oakforest-PACS}
\label{sec:KNL}

\subsection{Machine Environment}

We start with tuning on Oakforest-PACS.
We use the Intel compiler of version 18.0.1 with 
options \texttt{ -O3 -ipo -no-prec-div -xMIC-AVX512}.
On execution, job classes with the cache mode of MCDRAM are used.
According to the tuning-guide provided by JCAHPC,
we adopt the following setup.
To avoid OS jitters, the 0th and 1st cores on each KNL card
are not assigned for execution.
\texttt{KMP\_AFFINITY=compact} is set if multiple threads are assigned
to a core (unset for 1 thread/core).

\subsection{Matrix Multiplication}

\subsubsection{Vector Variables}

\begin{figure}[t]
\centering
 \includegraphics[width=0.52\linewidth,angle=-90]{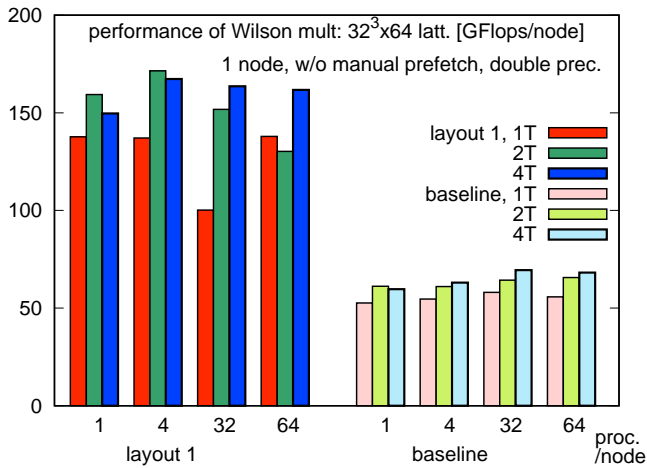}
\caption{
Effect of using the SIMD vector variables: the performance of
the Wilson matrix multiplication on a single KNL node.}
\label{fig:layout_dep_org_wilson}
\end{figure}

We first compare our new implementation with the core library of Bridge++,
which does not explicitly use the vector variables. 
Fig.~\ref{fig:layout_dep_org_wilson} shows the performance of the
Wilson matrix multiplication in double precision on a single node.
We use the layout 1 for the new code as it has similar site
layout on the memory.
The new code (layout 1) exhibits 2--3 times better performance
than the core library (baseline).
We also observe that the new code has a strong dependence on the number of
hyper-threading (denoted by 1T--4T).
As shown later, the best performance is achieved with 2T in most cases
for the new implementation.

\subsubsection{Data Layout}

\begin{figure}[t]
\centering
\includegraphics[width=0.52\linewidth,angle=-90]{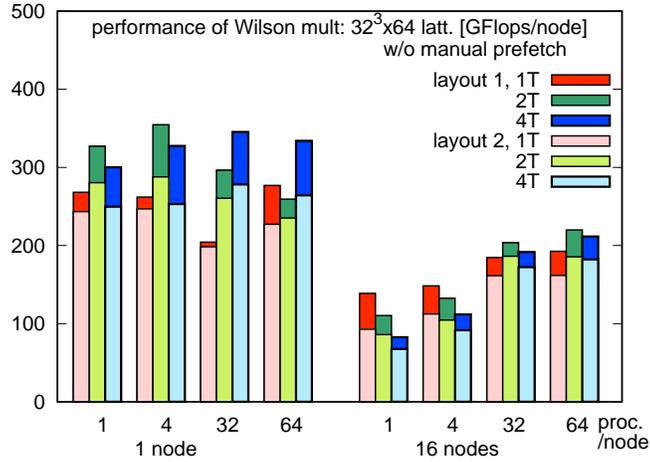}
\caption{
Data layout dependence: performance of the Wilson matrix multiplication
without prefetch on a single KNL node.}
\label{fig:layout_dep_wilson}
\end{figure}

Fig.~\ref{fig:layout_dep_wilson} compares the layout 1 and 2
for the Wilson matrix multiplication.
In this result the layout 1 is always faster.
A presumable reason is an extra shuffle
inside the SIMD vector for the layout 2 at the boundary of the local
lattice in $y$-, $z$- and $t$-directions.
In the current implementation of layout 2,
the data are always once shuffled even in the bulk and then
a mask operation chooses shuffled or unshuffled data.
Such a conditional shuffling does not exist in the layout 1.
Another possible reason is that, in packing the data for communication,
the layout 2 uses only the half of data in a SIMD vector.
Converting two SIMD vectors into one causes additional instructions and
additional load from the memory. 

As apparent in Fig.~\ref{fig:layout_dep_wilson} and successive two
figures, the performance depends on the number of hyper-threads/core
while no general tendency is observed.
Since it can be easily changed at run time, hereafter we focus on
the best performance case without indicating the number of
hyper-threads.

\subsubsection{Prefetching}
\begin{figure}[t]
\centering
 \includegraphics[width=0.52\linewidth,angle=-90]{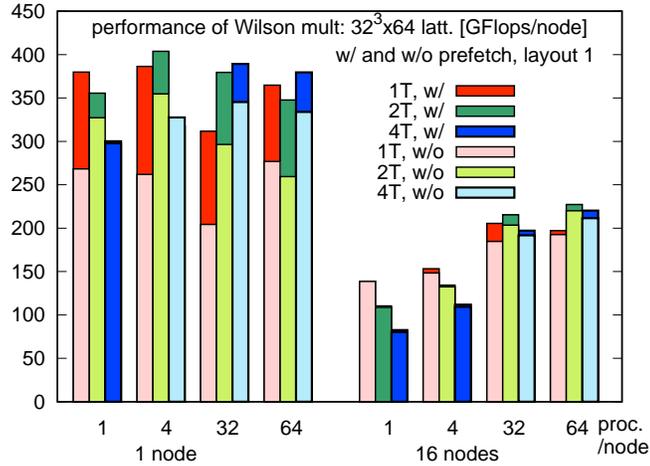}
\caption{
Effect of prefetch: performance of the Wilson matrix multiplication
with the layout 1 on a single KNL node.}
\label{fig:prefetch_dep_wilson}
\end{figure}

Effect of the manual prefetch against the automated prefetch by
compiler is displayed in
Fig.~\ref{fig:prefetch_dep_wilson}.
On a single node, where no inter-node communication is needed,
the manual prefetch achieves substantial improvement.
In the maximum performance case (4 MPI process) its effect is
more than 10\%, from 355 to 404 GFlops.
In some cases more gains are obtained.
Increasing the number of nodes, however, the effect is gradually
washed out and becomes only a few percent at 16 nodes in the
cases of 32 and 64 processes/node.
Sometimes the manual prefetch even slightly reduces the performance, for which
the colors are flipped in Fig.~\ref{fig:prefetch_dep_wilson}.
We observe a similar improvement for the clover matrix multiplication:
from 413 to 475 GFlops on single node (4 MPI proc./node)
and from 277 to 287 GFlops/node on 16 nodes (64 proc./node).

Since our target lattice sizes assume more than $O(10)$ KNL nodes,
the advantage of manual prefetch is not manifest compared to
involved tuning effort.
In the following, we nonetheless use the code with manual prefetches.
The performance without manual prefetch may be estimated based on
the result in Fig.~\ref{fig:prefetch_dep_wilson}.

For reference, here we quote the effect of communication
overhead for a single MPI process case on a single node.
As noted above, even in such a case the copy of packed boundary data
is performed.
By removing the redundant boundary data packing and copy,
the performance changes as follows:
378 $\rightarrow$ 453 GFlops (layout 1) and 
345 $\rightarrow$ 361 GFlops (layout 2) for the Wilson matrix,
and
430 $\rightarrow$ 497 GFlops (layout 1) and
388 $\rightarrow$ 440 GFlops (layout 2) for the clover matrix multiplication.

\subsubsection{Comparison to Other Codes}
Now we compare our performance of the Wilson and clover matrix
multiplication to other codes under the condition
of a single process on a single KNL node.
The QPhiX library achieves 587 GFlops for
single precision \cite{KNLtextbook} on a $32^3\times 96$ lattice.
The Grid library \cite{Boyle:2016lbp} provides a benchmark
of the Wilson matrix that we can run on the same environment as
this work.
On a $32^3\times 64$ lattice, based on v0.7.0,
it gives the best performance with one thread/core and amounts
to 340 GFlops that is comparable to our result.
According to Ref.~\cite{Rago:2017pyb}, the Grid achieves 960 GFlops 
with multiple right hand sides, that has an advantage in reuse of data. 
While our result is not as fast as QPhiX, it shows that
large fraction of performance can be achieved with rather simple
prescriptions.
An approach keeping the array of structure data layout and inserting
pragmas \cite{Durr:2017clx} gives 225 GFlops (245 GFlops
after correcting the difference in clock cycle).
In our previous report \cite{Kanamori:2017urm}, which corresponds to
the layout 2 without redundant boundary data packing/copy, the best performance
on single node was 340 GFlops (4 MPI proc./node).
With the same condition, it becomes 369 GFlops whose improvement is
mainly due to the refinement of the prefetch.
Boku {\it et al.} \cite{Boku:2017urp} reported that on the same machine
an even-odd preconditioned clover matrix multiplication,
which adopts different implementation from ours with the smaller
byte-per-flop value of 0.645, runs with about 560 GFlops/node up
to 8,000 KNL nodes.
The multi-node result with Grid reported in \cite{Boyle:2017xcy}
is 277 GFlops with a local lattice volume $24^4$.

\subsubsection{Scaling Property of Matrix Multiplication}

\begin{figure}[t]
\centering
\includegraphics[width=0.98\linewidth]{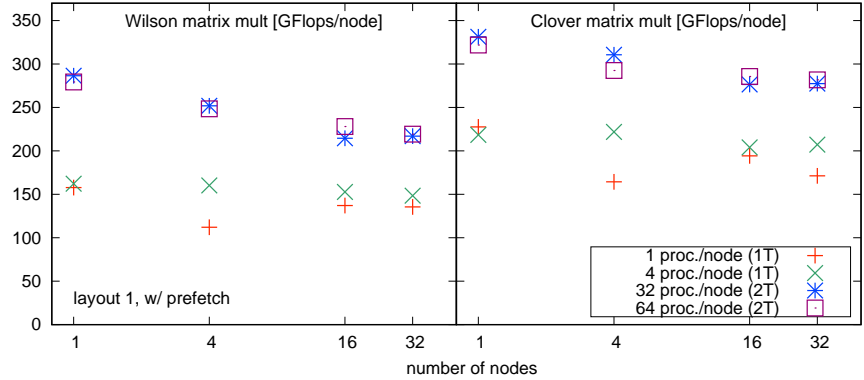}
\caption{
Weak scaling plots for the matrix multiplication 
with a $32\times 16^3$ local lattice in each node measured
on Oakforest-PACS.
}
\label{fig:mult_scaling}
\end{figure}

Fig.~\ref{fig:mult_scaling}
shows the weak scaling property up to 32 nodes for the Wilson (left) and the
clover (right) matrix multiplication with a local lattice volume 
$32\times 16^3$. 
The values are measured by averaging over successive 1,000 multiplications.
As expected from the byte-per-flop values, the clover matrix is more
efficient than the Wilson matrix.
For both the matrices, better performance is observed for 32 or 64 MPI processes
per node (1 process per tile or core) than the other two cases.
A similar tendency is observed in
Fig.~\ref{fig:prefetch_dep_wilson},
which corresponds to a strong scaling from 1 node to
16 nodes with a lattice volume $32^3\times 64$.

\subsection{Performance of BiCGStab Solver}

\begin{figure}[t]
\centering
\includegraphics[width=0.98\linewidth]{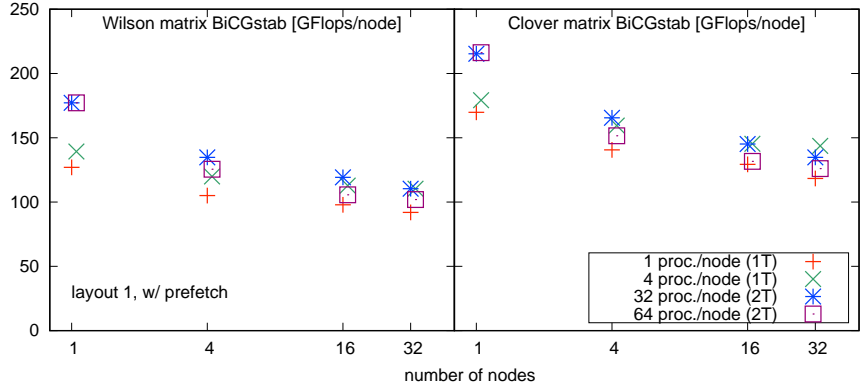}
\caption{
Weak scaling plots for the BiCGStab solver with the Wilson and clover
matrices with a $16^3 \times 32$ lattice in each node
measured on Oakforest-PACS.
Some symbols are slightly shifted horizontally to improve the visibility.
}
\label{fig:solver_scaling}
\end{figure}

For both the Wilson and clover matrices, the BiCGStab solver works efficiently.
We compose the solver algorithm with BLAS-like functions
to which neither manual prefetch nor additional compiler option for prefetch
is applied.
In Fig.~\ref{fig:solver_scaling},
we show the weak scaling plot of performance for the BiCGStab solver
with the Wilson and clover matrices.
The performance is an average over 12 times of solver call for
different source vectors.
Because of larger byte-per-flop values of the linear algebraic
functions, the performance reduces to about half the matrix
multiplication at 32 nodes%
\footnote{
The BiCGStab performance in the previous report \cite{Kanamori:2017urm}
was about the half of the actual one, due to a bug in counting 
the matrix multiplications.}.
Currently, each BLAS-like routine independently executes a load and store
of data.
Fusing several functions may improve the performance.

\section{Performance on Skylake-SP: ITO}
\label{eq:Skylake}

\subsubsection{Machine Environment}
Another machine we examine is ITO at Kyushu University.
The Intel compiler of version 18.0.0 is used
with the options \texttt{-ipo -O3 -no-prec-div -fp-mpdel fast=2 -xHost}.
At run time
an environment variable \texttt{KMP\_AFFINITY=compact} is set.

\subsubsection{Tuning}
We do not apply additional code tuning.
Actually, as shown below, the code tuned for KNL works reasonably
well on the Skylake-SP machines.
The effect of the manual prefetch, however, is quite limited or
even negative.
While refining prefetch tuning properly to the Skylake-SP processor
might improve the performance, we abandon to apply it.
We also observe that using hyper-threading
 --- up to 2 hyper-threads are possible --- spoils the performance
significantly, and thus do not use it.

\subsubsection{Matrix Multiplication}
\begin{figure}[t]
\centering
\includegraphics[width=0.98\linewidth]{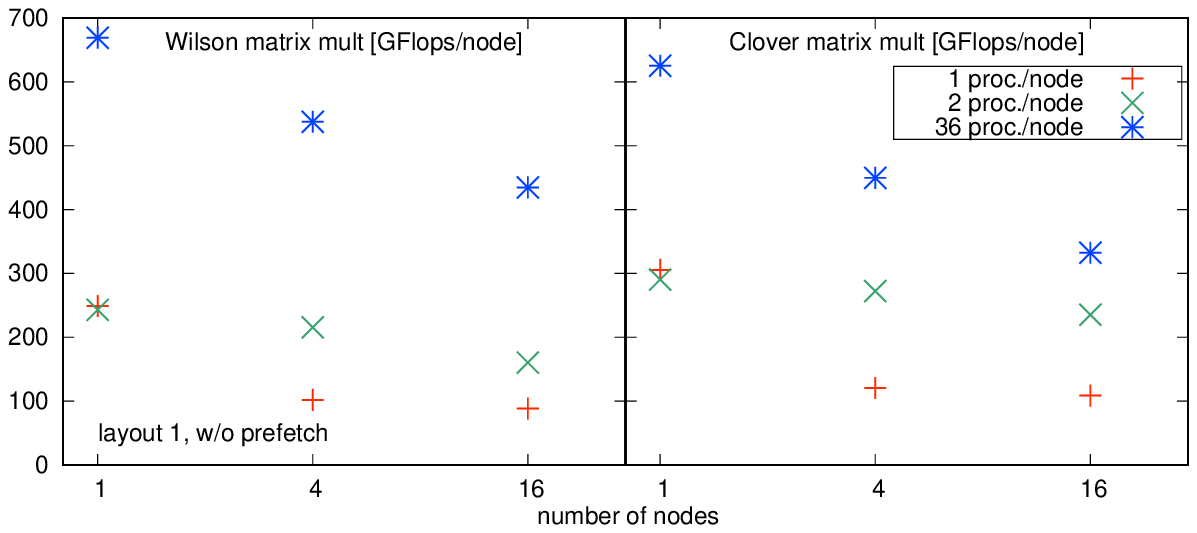}
\vspace{-0.2cm}\\
\includegraphics[width=0.98\linewidth]{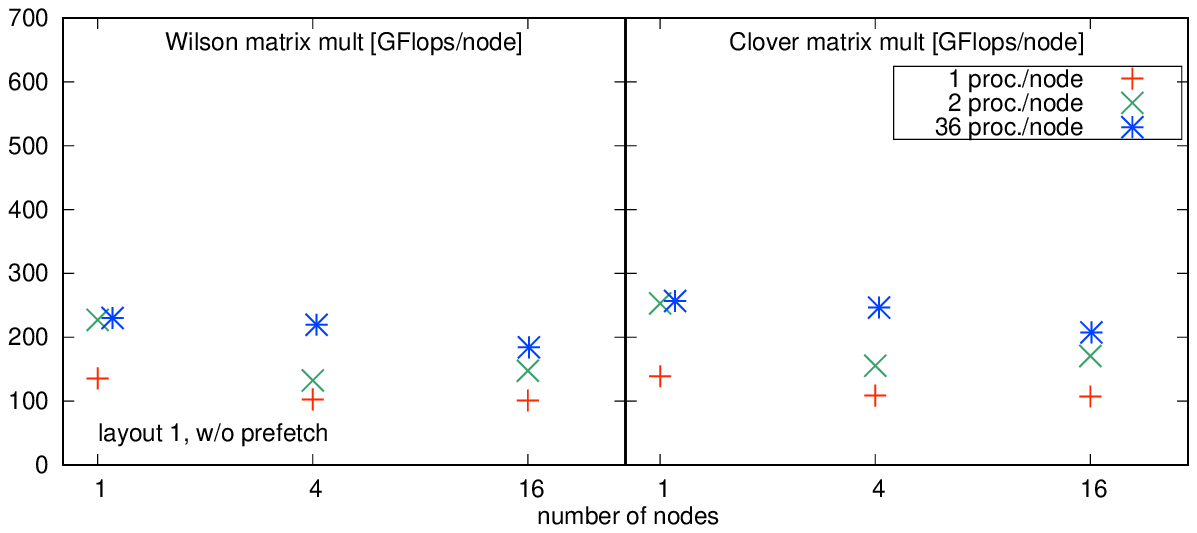}
\caption{
Weak scaling plots for the Wilson and clover matrix multiplication
on ITO.
The local lattice sizes on each node are $32 \times 16 \times 12 \times 12$
(top panel) and $64 \times 32 \times 24 \times 24$ (bottom).
}
\label{fig:ito_mult_scaling}
\end{figure}

Fig.~\ref{fig:ito_mult_scaling}
shows the weak scaling property of the matrix multiplication
up to 16 nodes.
In the top panel the lattice volume per node is
$32\times 16 \times 12 \times 12$, which is chosen to allow 36 MPI
processes per node.
We observe the performance with 36 MPI proc./node is much higher than the
other two cases.
The single node performance is 669 GFlops for the Wilson and
625 GFlops for the clover matrices, which is 91.8\% and 96.5\% of
the roofline limit of the ideal data reusing, respectively.
This implies that because of the small local lattice volume
reuse of data on cache is almost perfect.
As expected from the roofline estimate, the Wilson matrix exhibits
better performance than the clover matrix.
Increasing the lattice volume per node to $64\times 32^3$,
such high performance is lost
as shown in the bottom panel of Fig.~\ref{fig:ito_mult_scaling}.
The performance of the Wilson matrix multiplication is 230 GFlops,
which is slightly above the roof line estimation without any data reuse.
Note that the 2 proc./node result in the top panel of
Fig.~\ref{fig:ito_mult_scaling} is 243 GFlops.

Another observation is that the case with single MPI process per
node is the slowest.
Since in our implementation the communication is performed by
the master thread and other threads take charge of arithmetic
operations, the communication load becomes more significant
as the total number of threads per process increases.
In fact, for a single MPI process on single node, removing
redundant boundary data copy results in significant increase of
performance: 249 $\rightarrow$ 767 GFlops for the Wilson and 
306 $\rightarrow$ 624 GFlops for the clover matrices
on the $32\times 16 \times 12 \times 12$ lattice.
By replacing the MPI function with a substitution parallelized
with threads, the performance becomes about
80\% of that without copy of packed data.
This implies that one or a few process(es) per node is less
efficient as the local lattice size increases.
The performance of the Wilson matrix without boundary copy 
exceeds the roofline estimate with ideal reuse of cached data.
This may indicate that the data on the cache partially remain
until the next multiplication of the matrix.

\subsubsection{BiCGStab Solver}

\begin{figure}[t]
\centering
\includegraphics[width=0.95\linewidth]{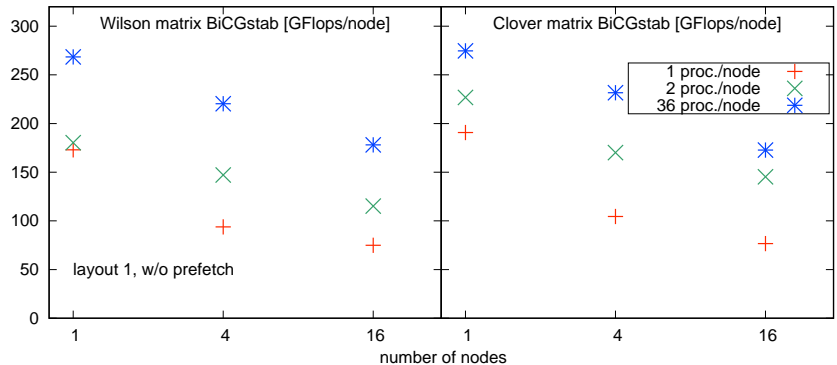}
\caption{
Weak scaling plots for the BiCGStab solver with the Wilson
and clover matrices with a $32 \times 16\times 12 \times 12$ local lattice
in each node measured on ITO.
}
\label{fig:ito_solver_scaling}
\end{figure}

Fig.~\ref{fig:ito_solver_scaling} shows the weak scaling of the
BiCGStab solver with a $32\times 16\times 12 \times 12$ local lattice
per node.
Again we observe high performance in the 36 proc./node case, which is
better than the result on single KNL node.
Because of higher requirement for the bandwidth in the linear algebra,
however, the performance is less than the matrix multiplication.
As observed in the matrix multiplication,
launching one MPI process per core achieves the best performance.

\section{Conclusion}

In this paper, we applied rather simple prescriptions to
make use of the SIMD architectures with the Intel AVX-512
instruction set to a typical problem in lattice QCD simulation.
Two different types of cluster systems were examined,
one composed of Intel Xeon Phi Knights Landing and the other of
Intel Xeon Skylake-SP.
We examine mainly the following prescriptions:
rearrangement of data layout, use of the AVX-512 intrinsics,
and manual prefetching.
The former two are crucial to achieve acceptable performance.
Not only to employ vector type variables, we compare two data
layouts and found the layout 1 achieves better performance
by about 10--20\%, while constraint on the MPI parallelization
is slightly stronger than the layout 2.
The effect of manual prefetching is more restrictive.
It is worth paying dedicated effort only on single or small
number of KNL nodes.

The same code is examined on a cluster system composed of Skylake-SP.
We observed that the code tuned for KNL exhibits a reasonable
performance, while the prefetch provides almost no effect.
If one chooses the lattice size and number of nodes appropriately
so that the problem size in each node is small enough, high
performance is expected by efficient reuse of cached data.
Since the L3 cache of Skylake-SP is faster than the MCDRAM of KNL,
it would be more effective to apply cache tuning, such as loop tiling.
For both the systems of KNL and Skylake-SP, we conclude that
running as a massive parallel machine is the efficient way.

\section*{Acknowledgment}

The authors would like to thank Peter Boyle, Guido Cossu,
Ken-Ichi Ishikawa, Daniel Richtmann, Tilo Wettig,
and the members of Bridge++ project for valuable discussion.
Numerical simulations were performed on the Oakforest-PACS system
hosted by JCAHPC, and the ITO system at 
Research Institute for Information Technology, Kyushu University.
We thank the support by Interdisciplinary Computational
Science Program in the Center for Computational Science (CCS),
University of Tsukuba, and by Intel Parallel Computing Center at CCS.
This work is supported by JSPS KAKENHI (Grant Numbers
JP25400284, JP16H03988), by
Priority Issue 9 to be tackled by Using Post K Computer,
and Joint Institute for Computational Fundamental Science (JICFuS).


\begin{thebibliography}{1}

 \bibitem{textbook}
	 For modern textbooks, {\it e.g.},
	 DeGrand, T., C.~DeTar, C.:
	 Lattice Methods for Quantum Chromodynamics.
	 World Scientific Pub., (2006);
	 Gattringer~C., Lang~C.~B.: Quantum Chromodynamics on the
	 Lattice. LNP, vol 788. Springer, Berlin Heidelberg (2010).
	 doi:10.1007/978-3-642-01850-3

\bibitem{Kanamori:2017tlp}
  Kanamori,~I., Matsufuru,~H.:
  Wilson and Domainwall Kernels on Oakforest-PACS.
  EPJ Web Conf.\  {\bf 175}, 09002 (2018).
  doi:10.1051/epjconf/201817509002
  [arXiv:1710.07226 [hep-lat]].
	
\bibitem{Kanamori:2017urm}
  Kanamori,~I., Matsufuru,~H.:
  Practical Implementation of Lattice QCD Simulation on Intel Xeon Phi
	Knights Landing.
	In: 2017 Fifth International Symposium on Computing and
	Networking (CANDAR), pp.~375--381, IEEE Explore (2018).
	doi:10.1109/CANDAR.2017.66
        [arXiv:1712.01505 [hep-lat]].

 \bibitem{KNLtextbook}
 Jeffers~J., Reinders,~J., Sodani~A.:
 Intel Xeon Phi Processor High Performance Programming Knights Landing
   Edition. Morgan Kaufmann, Cambridge (2016)

\bibitem{QPhiX}
 QPhiX library, \url{https://github.com/JeffersonLab/qphix}


\bibitem{Boyle:2016lbp} 
  Boyle,~P.~A., Cossu,~G., Yamaguchi,~A., Portelli,~A.:
  Grid: A next generation data parallel C++ QCD library.
  PoS LATTICE {\bf 2015}, 023 (2016).
  doi: 10.22323/1.251.0023
  [arXiv:1512.03487 [hep-lat]].

\bibitem{Boyle:2017xcy}
  Boyle,~P., Chuvelev,M., Cossu,~G., Kelly,~C., Lehner~,C. and Meadows,~L.:
  Accelerating HPC codes on Intel(R) Omni-Path Architecture networks: From particle physics to Machine Learning,
  arXiv:1711.04883 [cs.DC]

\bibitem{Boku:2017urp}
  Boku,~T., Ishikawa,~ K.-I., Kuramashi,~ Y., and Meadows,~L.:
  Mixed Precision Solver Scalable to 16000 MPI Processes for Lattice
	Quantum Chromodynamics Simulation on the Oakforest-PACS System.
	In: 2017 Fifth International Symposium on Computing and
	Networking (CANDAR), pp.~362--368 (2018), IEEE Explore (2018).
	doi:10.1109/CANDAR.2017.69
	[arXiv:1709.08785 [physics.comp-ph]].

\bibitem{Durr:2017clx} 
  Durr,~S.:
  Optimization of the Brillouin operator on the KNL architecture.
  EPJ Web Conf.\  {\bf 175}, 02001 (2018).
  doi:10.1051/epjconf/201817502001
  [arXiv:1709.01828 [hep-lat]].

\bibitem{Rago:2017pyb}
  Rago,~A.:
  Lattice QCD on new chips: a community summary
  EPJ Web Conf.\  {\bf 175}, 01021 (2018).
  doi:10.1051/epjconf/201817501021
  [arXiv:1711.01182 [hep-lat]].
	
 	
 \bibitem{KNLmemory}
	 Raman,~K. (Intel),
	 Optimizing Memory Bandwidth in Knights Landing on Stream Triad.
	\url{https://software.intel.com/en-us/articles/optimizing-memory-bandwidth-in-knights-landing-on-stream-triad}


\bibitem{bridge_website}
 Bridge++ project. \url{http://bridge.kek.jp/Lattice-code/}

\bibitem{Ueda:2014zsa}
 Ueda,~S. {\it et al.}:
 Bridge++: an object-oriented C++ code for lattice simulations.
	PoS LATTICE2013, 412 (2014).
	10.22323/1.187.0412.

\bibitem{Motoki:iccs2014}
 Motoki,~S. {\it et al.}:
	Development of Lattice QCD Simulation Code Set on Accelerators.
	Procedia Computer Science {\bf 29}, pp.~1701--1710 (2014).
	doi:10.1016/j.procs.2014.05.155;
%
 Matsufuru,~H. {\it et al.}: 
 OpenCL vs OpenACC: Lessons from Development of Lattice QCD
 Simulation Code.
 Procedia Computer Science {\bf 51}, pp.~1313-1322 (2015).
 doi:10.1016/j.procs.2015.05.316
	

\end{thebibliography}
\end{document}